\documentclass[aps,prl,twocolumn,groupedaddress]{revtex4-1}
\usepackage{amsmath}
\usepackage{graphicx}

\begin{document}


\title{Equivalence Principle and Bound Kinetic Energy}


\author{Michael A. Hohensee}
\email[]{hohensee@berkeley.edu}
\author{Holger M\"uller}
\affiliation{Department of Physics, University of California, Berkeley 94720, USA}

\author{R. B. Wiringa}
\affiliation{Physics Division, Argonne National Laboratory, Argonne, Illinois 60439, USA}


\date{\today}

\begin{abstract}

We consider the role of the internal kinetic energy of bound systems of matter in tests of the Einstein equivalence principle.  Using the gravitational sector of the standard model extension, we show that stringent limits on equivalence principle violations in antimatter can be indirectly obtained from tests using bound systems of normal matter.  We estimate the bound kinetic energy of nucleons in a range of light atomic species using Green's function Monte Carlo calculations, and for heavier species using a Woods-Saxon model.  We survey the sensitivities of existing and planned experimental tests of the equivalence principle, and report new constraints at the level of between a few parts in $10^{6}$ and parts in $10^{8}$ on violations of the equivalence principle for matter and antimatter.  

\end{abstract}


\maketitle

\newcommand{\ie}{\emph{i.e.}~}
\newcommand{\eg}{\emph{e.g.}~}
\newcommand{\cbw}{(\bar{c}^{w})}
\newcommand{\cbe}{(\bar{c}^{e})}
\newcommand{\cbp}{(\bar{c}^{p})}
\newcommand{\cbn}{(\bar{c}^{n})}
\newcommand{\abw}{(\bar{a}_{\rm eff}^{w})}
\newcommand{\abe}{(\bar{a}_{\rm eff}^{e})}
\newcommand{\abp}{(\bar{a}_{\rm eff}^{p})}
\newcommand{\abn}{(\bar{a}_{\rm eff}^{n})}
\newcommand{\cbTo}{(\bar{c}^{T}_{0})}
\newcommand{\cbT}{(\bar{c}^{T})}
\newcommand{\abT}{(\bar{a}_{\rm eff}^{T})}

General relativity follows from the Einstein equivalence principle (EEP), which holds that in any local Lorentz frame about any point in spacetime, the laws of physics are described by the standard model of particle physics and special relativity~\cite{MTW}.  Both general relativity and the standard model are believed to be the low energy limits of some as yet unknown complete theory of physics at high energy scales.  Such a theory might generate violations of EEP at experimentally accessible energy scales~\cite{Kostelecky:1989,Colladay:199798,Damour:1996}, although its exact form is unknown. Thus it is important to search for EEP violation in as many different places as possible.  
We use the Standard Model Extension (SME)~\cite{Colladay:199798}, a flexible and widely applied~\cite{datatables} framework for describing violations of EEP.  The SME is an effective field theory that phenomenologically augments the Standard Model action with terms that break local Lorentz invariance and other tenets of EEP~\cite{Kostelecky:2010}, while preserving energy conservation, gauge invariance, and general covariance.  As in other models~\cite{Damour:1996}, EEP-violation in the SME can manifest in multiple ways. In particular, it may be strongly suppressed in normal matter relative to antimatter~\cite{Kostelecky:2010,Hohensee:2011}.  Although the equivalence principle has been validated with extremely high precision for normal matter~\cite{adelberger}, the situation for antimatter is less clear.  

In this Letter, we show that in the SME, EEP violation in antimatter can be constrained by tests using bound systems of normal matter.  We clearly demonstrate how an anomaly that violates the weak equivalence principle for free particles generates anomalous gravitational redshifts in the energy of systems in which they are bound, in proportion to the systems' internal kinetic energy.  Using a nuclear shell model, we estimate the sensitivity of a variety of atomic nuclei to EEP violation for matter and antimatter, and illustrate points of commonality between older representations of EEP violation based on neutron excess and baryon number, and that of the SME.  We show that existing experimental~\cite{adelberger,Mueller:2010,matterwaves,Vessot:1980,PoundRebka,Kostelecky:2009a,Ashby, Blatt,Fortier,Hohensee:2013} limits on spin-independent EEP violation in matter and antimatter~\cite{Hohensee:2011} are up to ten times tighter than previously thought, and could be made tighter still, provided more precise estimates of the bound kinetic energy of particles in atomic systems.  
We focus on EEP-violation in conventional matter (made up of protons, neutrons, and electrons), and as in prior work~\cite{Kostelecky:2010,Hohensee:2011,datatables}, assume that anomalies affecting force-carrying virtual particles are negligible.  Using general covariance, we define our coordinates such that photons follow null geodesics, ensuring that electromagnetic fields do not violate EEP.
 
In the SME, spin-independent violations of EEP acting on a test particle of mass $m^{w}$ are described in its action~\cite{Kostelecky:2010}
\begin{equation}
S\!=\!-\!\!\!\int \!\frac{m^{w}c\sqrt{-(g_{\mu\nu}+2(\bar{c}^{w})_{\mu\nu})dx^{\mu}dx^{\nu}}}{1+(5/3)\cbw_{00}}+(a_{\rm eff}^{w})_{\mu}dx^{\mu},\!\!\label{eq:smeaction}
\end{equation}
where the superscript $w=p$, $n$, or $e$ (for proton, neutron, or electron) indicates the type of particle in question, $g_{\mu\nu}$ is the metric tensor, $dx^{\mu}$ is the interval between two points in spacetime, and $c$ is the speed of light.  The $\cbw_{\mu\nu}$ tensor describes a fixed background field that modifies the effective metric that the particle experiences, and thus its inertial mass relative to its gravitational mass.  The four-vector $(a^{w}_{\rm eff})_{\mu}=\{(1-U\alpha)\abw_{0},\abw_{j}\}$, where $U$ is the Newtonian potential, represents the particle's coupling to a field with a non-metric interaction with gravity.  As $(a^{w}_{\rm eff})_{\mu}$ is CPT-odd~\cite{Colladay:199798}, this term enters with opposite sign in the action for an antiparticle $\bar{w}$.  Both $\cbw_{\mu\nu}$ and $(a^{w}_{\rm eff})_{\mu}$ vanish if general relativity is valid.  For convenience, Eq.~\eqref{eq:smeaction} includes an unobservable scaling of the particle mass by $(1+\frac{5}{3}\cbw_{00})$.  

We focus on the isotropic subset of this model~\cite{Kostelecky:2010}, in which $\cbw_{\mu\nu}$ is diagonal and traceless, and the spatial terms in the vector $(a^{w}_{\rm eff})_{\mu}$ vanish.  In this limit, EEP-violation is described by the comparatively poorly constrained $\cbw_{00}$ coefficients~\cite{datatables}, and the $\abw_{0}$ terms, which cannot be measured in non-gravitational experiments~\cite{Colladay:199798}.  In the non-relativistic, Newtonian limit, less the rest mass energy and assuming that any violations of EEP must be small, the single particle Hamiltonian produced by the action \eqref{eq:smeaction} is given by
\begin{equation}
H=\frac{1}{2}m^{w}v^{2}-m^{w}_{g}U,\label{eq:nonrelfreeham}
\end{equation}
where the effective gravitational mass $m^{w}_{g}$ is given by\begin{equation*}
m^{w}_{g}=m^{w}\left(1-\frac{2}{3}(\bar{c}^{w})_{00}+\frac{2\alpha}{m^{w}}(\bar{a}^{w}_{\rm eff})_{0}\right).
\end{equation*}
Experimentally observable EEP violations are proportional to the particle's gravitational to inertial mass ratio
\begin{equation}\label{eq:betadef}
\frac{m_{g}^{w}}{m^{w}}=1-\frac{2}{3}\cbw_{00}+\frac{2\alpha}{m^{w}}\abw_{0}\equiv 1+\beta^{w},
\end{equation}
and are described here, as elsewhere~\cite{Mueller:2010,Hohensee:2011,Hohensee:2011a}, by the parameter $\beta^{w}$.  From Eq.~\eqref{eq:betadef}, we see that both $\cbw_{00}$ and $\abw_{0}$ are responsible for violations of the weak equivalence principle, an aspect of EEP~\cite{Will:2006}, since they produce particle-dependent rescalings of the effective gravitational potential.  In addition, EEP violation is not apparent in the non-relativistic motion of a free particle if $\alpha\abw_{0}=(m^{w}/3)\cbw_{00}$, although it remains manifest in the motion of the antiparticle $\bar{w}$, for which $\beta^{\bar{w}}=-2\alpha/m^{w}\abw_{0}-2/3\cbw_{00}$~\cite{Kostelecky:2010}.  As we now demonstrate, however, the antimatter anomaly $\beta^{\bar{w}}$ does contribute to tests involving non-gravitationally bound systems of matter, thanks to the anomalous gravitational redshift produced by $\cbw_{00}$ in the energies of bound systems.

For a bound system of particles, the total Hamiltonian is a sum of single-particle Hamiltonians, plus an interaction energy $V_{\rm int}$ that is assumed to be free of EEP-violating terms.  As implicit in Eq.~\eqref{eq:nonrelfreeham}, we take the system's squared center of mass velocity $\bar{v}^{2}$ to be small, and of similar order as the relevant change $\Delta U$ it explores in the gravitational potential.  Since the system is non-gravitationally bound, however, we cannot assume that the same is true of its constituent particles.  Thus we must include terms proportional to $v_{w,j}^{2}U/c^{2}$ in our Hamiltonian, where $v_{w,j}$ is the instantaneous velocity of the $j$th bound particle of species $w$.  In the limit that $\bar{v}\ll v_{w,j} \ll c$, we may approximate $v_{w,j}^{2}=(\bar{v}+\delta v_{w,j})^{2}\approx\bar{v}^{2}+(\delta v_{w,j})^{2}$, (dropping the mixed $\bar{v}(\delta v_{w,j})$ terms which make little contribution to the bound kinetic energy) and obtain
\begin{multline}
H=V_{\rm int}+\sum_{w}\Bigg[\frac{1}{2}m^{w}N^{w}\bar{v}^{2}-m^{w}N^{w}U(1+\beta^{w})\\
+\frac{1}{2}\sum_{j=1}^{N^{w}}(\delta v_{w,j})^{2}\left(1+\frac{3U}{c^{2}}+\frac{2U}{3c^{2}}\cbw_{00}\right)\Bigg].\label{eq:newrelham}
\end{multline}
The second line in Eq.~\eqref{eq:newrelham} represents the system's internally bound kinetic energy $T_{\rm int}$, and includes a term that contributes to the system's conventional gravitational redshift, as well as a term proportional to $\cbw_{00}$ and the gravitational potential $U$.  This last term corresponds to an anomalous gravitational redshift of the bound state energies.  To evaluate this term for bound quantum states, we recast it in terms of the momenta $\delta \vec{p}_{w,j}$ conjugate to the particle displacements $\delta x_{w,j}=x_{w,j}-\bar{x}$ from the system's center of mass $\bar{x}$.  The momenta satisfy $\delta \vec{p}_{w,j}=\partial H/\partial(\delta \vec{v}_{w,j})$, and so 
\begin{equation}
(\delta \vec{p}_{w,j})=m^{w}(\delta \vec{v}_{w,j})\left(1+\frac{3U}{c^{2}}+\frac{2U}{3c^{2}}\cbw_{00}\right).
\end{equation}
The bound kinetic energy $T_{\rm int}$ in Eq.~\eqref{eq:newrelham} is thus
\begin{equation}
T_{\rm int}=\sum_{w}\sum_{j=1}^{N^{w}}\frac{(\delta p_{w,j})^{2}}{2m^{w}}\left(1-\frac{3U}{c^{2}}-\frac{2U}{3c^{2}}\cbw_{00}\right).\label{eq:momexpr}
\end{equation}
Note that in general, to ensure that the system's mass defect is subject to a conventional gravitational redshift in the absence of EEP-violation, $V_{\rm int}$ must depend upon $U$.  If EEP is satisfied, the variation of the mass defect $m'_{A}=(V_{\rm int}+T_{\rm int})/c^{2}$ for a system $A$ in a gravitational potential $U$ is such that the ratio $m'_{A}(U_{1})/m'_{A}(U_{2})=1+(U_{1}-U_{2})/c^{2}$.  Due to our initial scaling of the particle mass in Eq.~\eqref{eq:smeaction}, the factor in parenthesis in Eq.~\eqref{eq:momexpr} contains terms proportional to $1$, $U$, $U\cbw_{00}$, but not $\cbw_{00}$ alone.  This, along with our assumption that $V_{\rm int}$ is independent of $\cbw_{00}$ and $\abw_{0}$, implies that the ratio $m'_{A}(U_{1})/m'_{A}(U_{2})$ does not generate additional cross terms in $U\cbw_{00}$, and we can therefore write the total Hamiltonian for a bound system $A$ as
\begin{equation}
H=\frac{1}{2}M_{A}\bar{v}^{2}-M_{A}U\left(1+\beta^{A}+\frac{2}{3}\sum_{w}\frac{T_{\rm int}^{w}}{M_{A}c^{2}}\cbw_{00}\right),\label{eq:kincorr}
\end{equation}
where $M_{A}=(\sum_{w}N^{w}m^{w})-m'_{A}$ incorporates the conventional components of $V_{\rm int}+T_{\rm int}$, the total kinetic energy of all $w$-particles in the system is {$T_{\rm int}^{w}=\nolinebreak\sum_{j=1}^{N^{w}}\langle(\delta p_{w,j})^{2}/2m^{w}\rangle$}, and
\begin{equation}\label{eq:betat}
\beta^{A}\equiv \frac{1}{M_{A}}\sum_{w}N^{w}m^{w}\left(\frac{2\alpha}{m^{w}}\abw_{0}-\frac{2}{3}\cbw_{00}\right).
\end{equation}
Since $\cbw_{00}=-(3/4)(\beta^{w}+\beta^{\bar{w}})$, this demonstrates that EEP tests using non-gravitationally bound systems of normal matter can constrain phenomena that would otherwise only be apparent for free antimatter particles.

We now apply Eq.~\eqref{eq:kincorr} to evaluate the phenomenological reach of existing experiments using conventional matter.  Violation of EEP is described by six independent parameters.  Three for matter: $\beta^{p}$, $\beta^{n}$, and $\beta^{e}$; and three for antimatter: $\beta^{\bar{p}}$, $\beta^{\bar{n}}$, and $\beta^{\bar{e}}$.  For any particular EEP test comparing the effects of gravity acting on systems $A$ and $B$, the observable anomaly is given by $\beta^{A}-\beta^{B}$, where $\beta^{A}$ and $\beta^{B}$ are defined in Eqs.~\eqref{eq:kincorr} and~\eqref{eq:betat}.  Since all high-precision tests of EEP are performed on charge-neutral systems, and since normal matter has a substantially similar ratio of proton to neutron content, the expression for $\beta^{A}-\beta^{B}$ can be usefully expressed in terms of an effective neutron excess $\widetilde{\Delta}_{j}$, effective mass defect $\widetilde{m}'_{j}$, and kinetic energy components $T^{w}_{j,{\rm int}}$ of the two systems, where 
\begin{eqnarray}
\widetilde{\Delta}_{j}&\equiv&\frac{m^{n}}{m^{p}}\frac{m^{e}+m^{p}}{m^{n}}N_{j}^{n}-N_{j}^{p},\\
\widetilde{m}'_{j}&\equiv&m'_{j}-\frac{(m^{n}-m^{p})(m^{e}+m^{p})}{m^{n}}N_{j}^{p},
\end{eqnarray}
and $j\in\{A,B\}$.  
The EEP-violating observable can then be written in terms of linear combinations of the free particle ($\beta^{w}$) and anti-particle ($\beta^{\bar{w}}$) anomalies as
\begin{multline}
\beta^{A}-\beta^{B}=\frac{(m^{n})^{2}}{(m^{n})^{2}+(m^{e}+m^{p})^{2}}\!\!\left[\phantom{\left(\frac{\widetilde{\Delta}_{A}}{M_{A}}-\frac{\widetilde{\Delta}_{B}}{M_{B}}\right)}\right.\\
\left.\left(\frac{\widetilde{\Delta}_{A}}{M_{A}}-\frac{\widetilde{\Delta}_{B}}{M_{B}}\right)m^{p}\beta^{e+p-n}-\left(\frac{\widetilde{m}'_{A}}{M_{A}}-\frac{\widetilde{m}'_{B}}{M_{B}}\right)\beta^{e+p+n}\right]\\
-\frac{1}{2}\sum_{w}\left(\frac{T^{w}_{A, {\rm int}}}{M^{A}c^{2}}-\frac{T^{w}_{B, {\rm int}}}{M^{B}c^{2}}\right)\left(\beta^{w}+\beta^{\bar{w}}\right)\label{eq:betadiffdefect},
\end{multline}
where $M_{A}$ and $M_{B}$ are the masses of the two test bodies, and 
\begin{eqnarray}
\beta^{e+p-n}&\equiv&\beta^{e+p}-\frac{m^{e}+m^{p}}{m^{n}}\beta^{n}\label{eq:betaeppmn}\\
\beta^{e+p+n}&\equiv&\frac{m^{e}+m^{p}}{m^{n}}\beta^{e+p}+\beta^{n},\label{eq:betaepppn}
\end{eqnarray}
in which 
\begin{equation}
\beta^{e+p}\equiv\frac{m^{e}}{m^{p}}\beta^{e}+\beta^{p}\label{eq:betaepp},
\end{equation}
after the notation of~\cite{Kostelecky:2010}.  We can define a similar set of terms $\beta^{\bar{e}+\bar{p}}$, $\beta^{\bar{e}+\bar{p}-\bar{n}}$, and $\beta^{\bar{e}+\bar{p}+\bar{n}}$ for antimatter.  Note that Eq.~\eqref{eq:betadiffdefect} has a close parallel with older studies of EEP-violation~\cite{Damour:1996}, since 
\begin{equation}
\left(\frac{\widetilde{m}'_{B}}{M_{B}}-\frac{\widetilde{m}'_{A}}{M_{A}}\right)=\left(\frac{\widetilde{A}_{B}}{M_{B}}-\frac{\widetilde{A}_{A}}{M_{A}}\right)m^{n},
\end{equation}
where the effective baryon number $\widetilde{A}_{j}$ is given by
\begin{equation}
\widetilde{A}_{j}\equiv N_{j}^{n}+\frac{m^{p}}{m^{n}}\frac{m^{e}+m^{p}}{m^{n}}N_{j}^{p}.
\end{equation}
Thus the quantities $m^{p}\beta^{e+p-n}$ and $m^{n}\beta^{e+p+n}$ in the SME may be understood as parameterizing an anomalous gravitational coupling to a given particle's neutron-excess and total baryon number ``charges''~\cite{Damour:1996}.

 \begin{table}
 \caption{\label{tab:gfmckine}Comparison between calculated bound kinetic energies (in MeV) of protons and neutrons in light nuclei, obtained from many-body Green's function Monte-Carlo (GFMC) calculations~\cite{Pieper:2001}, and a single-particle calculation using a modified Woods-Saxon potential.}
 \begin{ruledtabular}
 \begin{tabular}{lrrrr}
Species & \multicolumn{2}{c}{GFMC} & \multicolumn{2}{c}{Woods-Saxon}\\
& $T_{\rm int}^{\rm p}$ & $T_{\rm int}^{\rm n}$ & $T_{\rm int}^{\rm p}$ & $T_{\rm int}^{\rm n} $\\
$^{6}$Li & 77  & 78  & 64  & 65 \\
$^{7}$Li & 88  & 108  & 67  & 84 \\
$^{9}$Be & 124  & 135  & 95  & 112 \\
$^{10}$B & 162  & 164  & 116  & 122 \\
$^{12}$C & 219  & 219  & 145  & 153 \\
 \end{tabular}
 \end{ruledtabular}
 \end{table}

\begin{figure}[tp]
\includegraphics[width=3.4in]{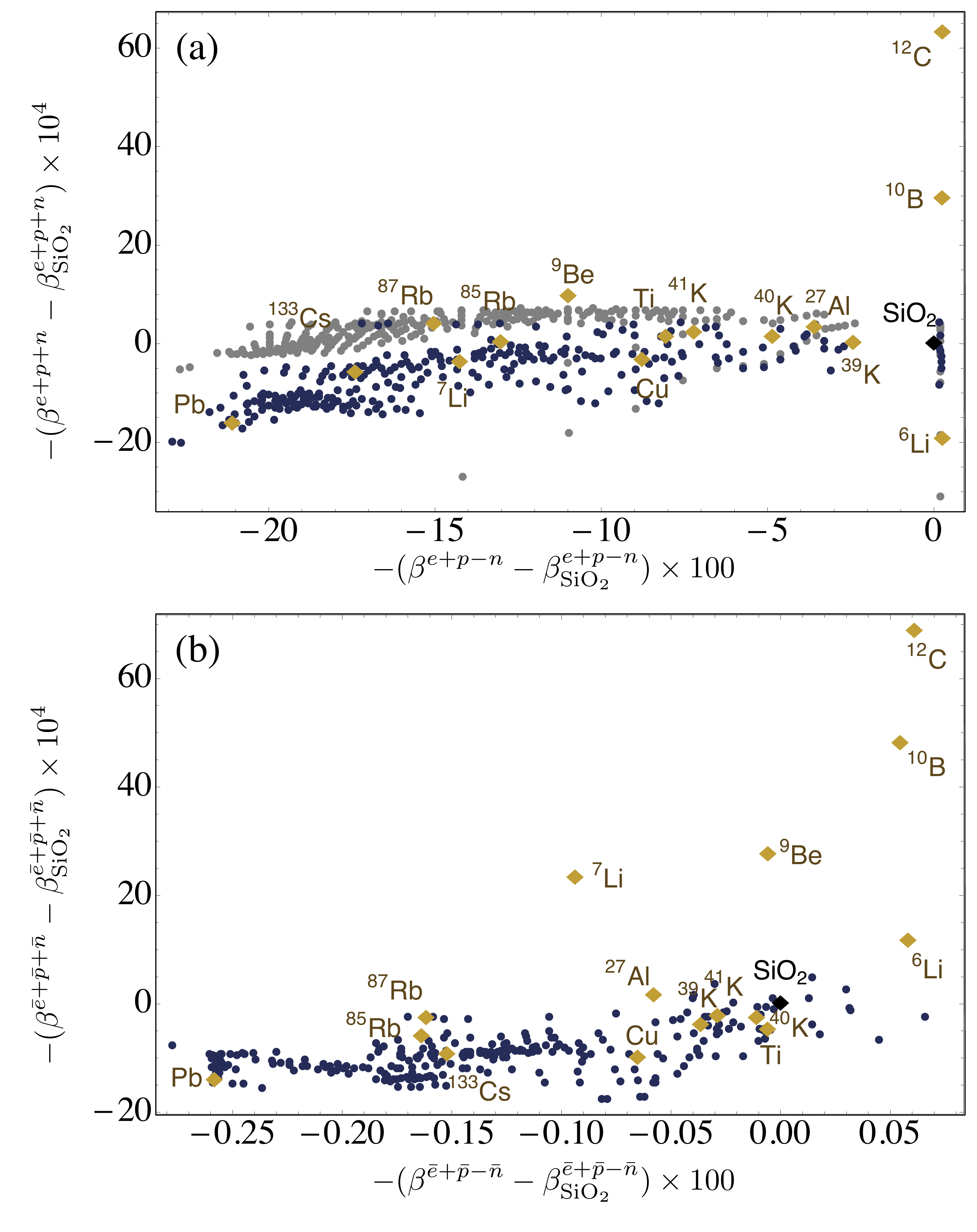}
\caption{\label{fig:scatterplot} Scatterplot of the contribution of $\beta^{e+p\pm n}$ and $\beta^{\bar{e}+\bar{p}\pm\bar{n}}$ parameters to observable EEP violation in normal nuclides with lifetimes in excess of 1 Gyr, when compared to SiO$_{2}$.  Tests that compare two or more widely separated species are more sensitive than tests involving neighboring isotopes.  Plot (a) shows each species' relative sensitivity to matter-sector EEP-violation, and (b) depicts their sensitivities to antimatter-sector anomalies.  Gray points in (a) indicate the range of sensitivities obtained without accounting for nucleons' kinetic energies.  Sensitivities of $^{6}$Li, $^{7}$Li, $^{9}$Be, $^{10}$B, and $^{12}$C are taken from GFMC calculations, all others from a Woods-Saxon model (see Supplement~\cite{supplement}).}
\end{figure}

In our prior analysis~\cite{Hohensee:2011}, the kinetic energy of protons and neutrons bound within a given nucleus was estimated by treating the nucleons as Fermi gases confined within a square potential well.  This model did not account for the nucleons' angular momentum, treated the Coulomb potential in a heuristic way by shifting the depth of the proton potential, and did not account for the nucleons' spin-orbit interaction.  The latter is of particular significance, because it can affect the occupation number of states with a given kinetic energy.  Here, we improve upon that work by modeling the nucleons as single particles bound within fixed, spherically symmetric rounded square well potentials.  These Woods-Saxon potentials~\cite{Woods:1954} are taken to be of the form developed by Schwierz \emph{et al.}~\cite{Schwierz:2007}.  Nuclide data is taken from Audi \emph{et al.}~\cite{Audi:2003}, and isotopic abundances (for deriving the EEP-violating signal in bulk materials) from Laeter \emph{et al.}~\cite{Laeter:2003}.  A complete summary of our calculated kinetic energies can be found in the Supplement~\cite{supplement}.  Better estimates of the nucleons' bound kinetic energies are available for light nuclei using Green's function Monte-Carlo (GFMC) calculations of the many-nucleon wave functions for nuclides with $A\leq 12$~\cite{Pieper:2001}.  The GFMC estimates of the bound kinetic energy of the constituent protons and neutrons in $^{6}$Li, $^{7}$Li, $^{9}$Be, $^{10}$B, and $^{12}$C are summarized in Tab.~\ref{tab:gfmckine}, and are compared with the corresponding predictions of our Woods-Saxon potential.  Using these estimates, we can determine the contribution of the matter-sector $\beta^{e+p\pm n}$ and antimatter-sector $\beta^{\bar{e}+\bar{p}\pm\bar{n}}$ parameters to any observed violation of EEP in the motion of two (normal matter) test masses.  These contributions are summarized in Fig.~\ref{fig:scatterplot}.  Species with particular relevance to existing or planned tests of EEP~\cite{Dimopoulos,DropTower,kasevich,STEQuest,SRPoem,lithium,GG} are explicitly labeled.

In most experiments, $\beta^{e+p-n}$ is dominant, as it scales with the neutron excess.  The next most accessible are the $\beta^{e+p+n}$ term, which scales with the mass defect, and the antimatter term $\beta^{\bar{e}+\bar{p}-\bar{n}}$, which scales with the excess of the neutrons' kinetic energy over that of the protons, followed by $\beta^{\bar{e}+\bar{p}+\bar{n}}$.  In some cases, (\eg tests comparing lead and aluminium~\cite{SRPoem}) the signal from the antimatter $\beta^{\bar{e}+\bar{p}-\bar{n}}$ may actually be stronger than that from $\beta^{e+p+n}$.  These terms represent four of the six degrees of freedom describing isotropic EEP violation, primarily for protons, neutrons and their antiparticles.  Electronic EEP-violation is described by $\beta^{e-p}+\beta^{\bar{e}-\bar{p}}\equiv -\frac{4}{3}[\cbe_{00}-\frac{m^{e}}{m^{p}}\cbp_{00}]$, and has thus far been constrained largely by gravitational redshift tests~\cite{Vessot:1980,PoundRebka,Ashby,Blatt,Fortier,Hohensee:2013}, and tests of local Lorentz invariance~\cite{Kostelecky:1999,AltschulLehnert}.  The sixth degree of freedom, $\beta^{e-p}-\beta^{\bar{e}-\bar{p}}\propto \alpha\abp_{0}-\alpha\abe_{0}$, is only observable in tests on charged bodies~\cite{Kostelecky:2010,Hohensee:2011a}.

Using multivariate normal analysis of the results of an ensemble of EEP tests, including matter-wave~\cite{Hohensee:2011,Mueller:2010,matterwaves}, clock comparison~\cite{PoundRebka,Vessot:1980,Kostelecky:2009a,Ashby,Blatt,Fortier,Hohensee:2013}, and torsion pendulum experiments~\cite{adelberger}, we obtain new limits on the five isotropic EEP-violating degrees of freedom that are observable in neutral systems, summarized in Tab.~\ref{tab:newlimits}.  These bounds improve upon prior~\cite{Hohensee:2011} gravitational constraints on these SME coefficients by factors of two to ten, and are also stated in terms of the five matter and anti-matter $\beta^{e+p\pm n}$, $\beta^{\bar{e}+\bar{p}\pm\bar{n}}$, and $\beta^{e-p}+\beta^{\bar{e}-\bar{p}}$ coefficients.  Though the limits reported in Tab.~\ref{tab:newlimits} are necessarily model-dependent, they are stable against small variations in the estimated value of $T^{w}/Mc^{2}$ for the relevant nuclides, and are consistent with the limits obtained using substantially different nuclear models~\cite{WitekTBP}.
\begin{table}[t]
\caption{\label{tab:newlimits}Global limits ($\times 10^{6}$) on isotropic EEP-violation, obtained via multivariate normal analysis on the results of an ensemble of precision tests of EEP.  Limits are stated in the Sun-Centered, Celestial Equatorial Frame~\cite{datatables}, and are expressed in terms of the $\beta^{w}$ parameters as well as the individual $(\bar{c}^{w})_{TT}$ and $\alpha(\bar{a}^{w})_{T}$, with $(\bar{a}^{e+p})_{T}\equiv(\bar{a}^{e})_{T}+(\bar{a}^{p})_{T}$.  Also shown is the limit on the $1\sigma$ volume $\beta^{\Pi}$ of five-dimensional parameter space consistent with experiment. }
\begin{tabular}{lr|lr}\hline\hline
$(\beta^{e-p}+\beta^{\bar{e}-\bar{p}})$ & $0.019\pm0.037$ & $(\bar{c}^{e})_{TT}$ & $-0.014\pm0.028$\\
$\beta^{e+p-n}$ & $-0.013\pm0.021$ & $(\bar{c}^{n})_{TT}$ & $1.1\pm1.4$\\
 $\beta^{e+p+n}$ &  $2.4\pm3.9$ & $(\bar{c}^{p})_{TT}$ & $0.24\pm0.30$\\
  $\beta^{\bar{e}+\bar{p}-\bar{n}}$ & $1.1\pm1.8$ & $\alpha(\bar{a}^{n})_{T}$ & $0.51\pm0.64$\\
   $\beta^{\bar{e}+\bar{p}+\bar{n}}$ & $-4.1\pm6.7$ & $\alpha(\bar{a}^{e+p})_{T}$ & $0.22\pm0.28$\\
\hline\hline
\end{tabular}
\end{table}

Despite the fact that torsion pendulum tests~\cite{adelberger} set limits on specific combinations of $\beta$ parameters at the level of $10^{-12}$ (having constrained $\Delta g/g$ to the level of $10^{-14}$), the best bounds reported in Tab.~\ref{tab:newlimits} are at the level of $10^{-8}$.  This apparent discrepancy is due to the fact that such tests do not span the full parameter space considered here.  Thus the limits on the individual $\beta$'s summarized in Tab.~\ref{tab:newlimits} are strongly correlated with one another.  Analysis of these correlations reveals that some combinations of the $\beta$'s are indeed constrained at the level of $10^{-9}$, $10^{-11}$ and $10^{-12}$, thanks to matter-wave interferometer and torsion pendulum results.  Unfortunately, the specific combinations of $\beta$'s subject to these constraints are sensitive to small errors in our estimates of the nuclides' bound kinetic energy, due to disparities between the precision of torsion pendulums and of other EEP tests.  Formal limits on EEP-violation at the level of an effective field theory like the SME must therefore await the development of more reliable nuclear models~\cite{WitekTBP} or the results of additional high precision EEP test presently in development, using matter-waves~\cite{DropTower,lithium,kasevich}, clocks~\cite{STEQuest} or macroscopic masses~\cite{SRPoem,GG}.

We have demonstrated that EEP tests on non-gravitationally bound systems of normal particles can set indirect constraints on EEP-violation in antimatter, thanks to the interaction between the EEP-violating terms and the system's bound kinetic energy.  We have explicitly derived the link between anomalous gravitational redshifts and violations of the weak equivalence principle.  This occurs whenever EEP is violated by introducing a particle-specific metric.  In the context of the SME, accounting for these interactions results in significantly improved constraints on EEP-violation in the standard model lagrangian, for both matter and antimatter.  The precision of these bounds is limited by that of existing nuclear models, and uneven experimental coverage of EEP-violating parameter space.  New EEP tests with precision comparable to that of existing torsion pendulum experiments~\cite{DropTower,lithium,kasevich,STEQuest,SRPoem,GG} may substantially eliminate this model-dependent limitation.  Better nuclear modeling could also improve limits on EEP violation in the SME by up to eight orders of magnitude, the pursuit of which will be the subject of future work.

We thank Brian Estey, Paul Hamilton, Alan Kosteleck\'y and Jay Tasson for stimulating discussions.  We also thank W. Nazarewicz and N. Birge for providing us with independent estimates of the bound kinetic energy of nucleons in a range of atomic species.

\section*{Supplemental Material}
In a previous analysis~\cite{Hohensee:2011}, we estimated the kinetic energy of the protons and neutrons bound within a given nucleus by treating the nucleons as fermi gases confined to a square potential well.  Here we improve upon that work using a shell model calculation.  The nucleus is modeled as a pair of rounded, spherically symmetric square well, or Woods-Saxon, potentials which separately confine its constituent protons and neutrons.  Our potential is that of~\cite{Schwierz:2007}, although we do not work with relative coordinates, and so do not use the reduced particle mass in our Hamiltonian.  For a nucleon of mass $m^{w}$ in a nucleus with mass number $A=Z+N$, made up of $Z$ protons and $N$ neutrons, our model Hamiltonian is~\cite{Schwierz:2007}
\begin{multline}
H=\frac{p^{2}}{2m^{w}}+V_{0}\left(1-\frac{4\kappa}{A}\langle \mathbf{t}\cdot\mathbf{T'}\rangle\right)f(r,R,a)+V_{c}(r,R)\\
+\frac{1}{2(m^{w})^{2}r}\left(\frac{\partial}{\partial r}\tilde{V}f(r,R_{SO},a)\right)\mathbf{L}\cdot\mathbf{S},\label{eq:WSHam}.
\end{multline}
The Woods-Saxon potential is given by
\begin{equation}
f(r,R,a) = \frac{1}{1+e^{(r-R)/a}},
\end{equation}
and $V_{0}=-52.06$ MeV, $\tilde{V}=24.1V_{0}$, $R=1.26 A^{1/3}$ fm, $R_{SO}=1.16 A^{1/3}$ fm, $a=0.662$ fm.  The vectors $\mathbf{t}$ and $\mathbf{T'}$ are the isospin of the nucleon and of the nucleus less that nucleon, respectively, and as in~\cite{Schwierz:2007}, are taken to be such that
\begin{equation}
-4\langle \mathbf{t}\cdot\mathbf{T'}\rangle=\left\{\begin{aligned} 
&3, && N=Z\\
&\pm(N-Z+1)+2, && N>Z\\
&\pm(N-Z-1)+2, && N<Z,
\end{aligned}\right.
\end{equation}
with $\kappa=0.639$.  The Coulomb potential $V_{c}(r)$ applies only to protons, and is given by
\begin{equation}
V_{c}(r,R)=(Z-1)e^{2}\left\{\begin{aligned} &\frac{3R^{2}-r^{2}}{2R^{3}} , && r\leq R\\ &\frac{1}{r}, && r> R.\end{aligned}\right.
\end{equation}

We solve for the eigenstates of this Hamiltonian numerically, and assign the protons and neutrons respectively to the $Z$ and $N$ lowest-lying energy states.  We then evaluate and sum the expectation value of the kinetic energy operator $\langle p^{2}/2m\rangle$ for each occupied state, to obtain the kinetic energy correction term of Eq.~(7).  The total estimated nucleon kinetic energies for all stable, and many long-lived nuclides are shown in Tables~\ref{tab:kinenergies} and~\ref{tab:kinenergieslonglived}.  The accuracy of these estimates is not guaranteed, as experimental measurements of the nucleons' bound kinetic energy are unavailable, and these results are derived from models that have been optimized for the solution of other problems~\cite{Schwierz:2007}.  Nevertheless, they do yield limits on the $\beta$ coefficients (see Tab.~II) that are consistent with those derived using other nuclear models~\cite{WitekTBP}, and exhibit similar trends.


\begin{table*}[ht]
\caption{\label{tab:kinenergies}Estimated bound kinetic energies (in MeV) of protons ($T^{p}_{\rm int}$) and neutrons ($T^{n}_{\rm int}$) in stable nuclides.}
\begin{tabular}{lrr|lrr|lrr|lrr|lrr}\hline\hline
Species & $T_{\rm int}^{p}$ & $T_{\rm int}^{n}$ &  Species & $T_{\rm int}^{p}$ & $T_{\rm int}^{n}$ & Species & $T_{\rm int}^{p}$ & $T_{\rm int}^{n}$ & Species & $T_{\rm int}^{p}$ & $T_{\rm int}^{n}$ & Species & $T_{\rm int}^{p}$ & $T_{\rm int}^{n}$ \\
$^{6}$Li & 64 & 65 & $^{54}$Cr & 559 & 659 & $^{94}$Mo & 831 & 1079 & $^{130}$Ba & 1211 & 1636 & $^{169}$Tm & 1405 & 2233 \\

$^{7}$Li & 67 & 84 & $^{54}$Fe & 590 & 662 & $^{95}$Mo & 886 & 1097 & $^{131}$Xe & 1173 & 1692 & $^{170}$Er & 1383 & 2266 \\

$^{9}$Be & 95 & 112 & $^{55}$Mn & 576 & 667 & $^{96}$Mo & 903 & 1114 & $^{132}$Xe & 1171 & 1712 & $^{170}$Yb & 1424 & 2157 \\

$^{10}$B & 116 & 122 & $^{56}$Fe & 574 & 674 & $^{96}$Ru & 869 & 1088 & $^{132}$Ba & 1208 & 1678 & $^{171}$Yb & 1422 & 2188 \\

$^{11}$B & 124 & 143 & $^{57}$Fe & 575 & 680 & $^{97}$Mo & 903 & 1131 & $^{133}$Cs & 1189 & 1716 & $^{172}$Yb & 1421 & 2311 \\

$^{12}$C & 145 & 153 & $^{58}$Fe & 576 & 686 & $^{98}$Mo & 903 & 1148 & $^{134}$Xe & 1168 & 1753 & $^{173}$Yb & 1419 & 2249 \\

$^{13}$C & 154 & 165 & $^{58}$Ni & 608 & 685 & $^{98}$Ru & 868 & 1124 & $^{134}$Ba & 1206 & 1719 & $^{174}$Yb & 1417 & 2279 \\

$^{14}$N & 165 & 173 & $^{59}$Co & 593 & 692 & $^{99}$Ru & 924 & 1141 & $^{135}$Ba & 1204 & 1739 & $^{175}$Lu & 1436 & 2295 \\

$^{15}$N & 172 & 185 & $^{60}$Ni & 610 & 697 & $^{100}$Ru & 958 & 1159 & $^{136}$Xe & 1165 & 1793 & $^{176}$Yb & 1413 & 2340 \\

$^{16}$O & 183 & 191 & $^{61}$Ni & 610 & 721 & $^{101}$Ru & 958 & 1197 & $^{136}$Ba & 1203 & 1760 & $^{176}$Hf & 1454 & 2308 \\

$^{17}$O & 188 & 211 & $^{62}$Ni & 611 & 745 & $^{102}$Ru & 958 & 1217 & $^{136}$Ce & 1239 & 1724 & $^{177}$Hf & 1452 & 2338 \\

$^{18}$O & 192 & 215 & $^{63}$Cu & 616 & 751 & $^{102}$Pd & 904 & 1224 & $^{137}$Ba & 1201 & 1780 & $^{178}$Hf & 1451 & 2368 \\

$^{19}$O & 211 & 235 & $^{64}$Ni & 610 & 792 & $^{103}$Rh & 985 & 1268 & $^{138}$Ba & 1200 & 1799 & $^{179}$Hf & 1449 & 2398 \\

$^{20}$Ne & 229 & 240 & $^{64}$Zn & 622 & 776 & $^{104}$Ru & 957 & 1254 & $^{138}$Ce & 1236 & 1765 & $^{180}$Hf & 1447 & 2428 \\

$^{21}$Ne & 233 & 258 & $^{65}$Cu & 616 & 799 & $^{104}$Pd & 1011 & 1279 & $^{139}$La & 1217 & 1802 & $^{180}$W & 1487 & 2390 \\

$^{22}$Ne & 236 & 264 & $^{66}$Zn & 622 & 825 & $^{105}$Pd & 1011 & 1306 & $^{140}$Ce & 1233 & 1805 & $^{181}$Ta & 1465 & 2439 \\

$^{23}$Na & 254 & 289 & $^{67}$Zn & 622 & 849 & $^{106}$Pd & 1011 & 1333 & $^{141}$Pr & 1241 & 1807 & $^{182}$W & 1483 & 2449 \\

$^{24}$Mg & 271 & 300 & $^{68}$Zn & 622 & 872 & $^{106}$Cd & 996 & 1295 & $^{142}$Ce & 1230 & 1840 & $^{183}$W & 1481 & 2479 \\

$^{25}$Mg & 274 & 321 & $^{69}$Ga & 628 & 878 & $^{107}$Ag & 1038 & 1342 & $^{142}$Nd & 1249 & 1808 & $^{184}$W & 1479 & 2508 \\

$^{26}$Mg & 276 & 319 & $^{70}$Zn & 638 & 882 & $^{108}$Pd & 1010 & 1386 & $^{143}$Nd & 1248 & 1826 & $^{184}$Os & 1520 & 2517 \\

$^{27}$Al & 292 & 352 & $^{70}$Ge & 634 & 883 & $^{108}$Cd & 1064 & 1349 & $^{144}$Sm & 1265 & 1811 & $^{185}$Re & 1497 & 2550 \\

$^{28}$Si & 308 & 361 & $^{71}$Ga & 627 & 888 & $^{109}$Ag & 1037 & 1395 & $^{145}$Nd & 1245 & 1862 & $^{186}$W & 1475 & 2648 \\

$^{29}$Si & 310 & 361 & $^{72}$Ge & 633 & 894 & $^{110}$Pd & 1009 & 1438 & $^{146}$Nd & 1243 & 1880 & $^{187}$Os & 1513 & 2581 \\

$^{30}$Si & 349 & 346 & $^{73}$Ge & 668 & 920 & $^{110}$Cd & 1063 & 1402 & $^{148}$Nd & 1258 & 1915 & $^{188}$Os & 1511 & 2603 \\

$^{31}$P & 312 & 369 & $^{74}$Ge & 668 & 947 & $^{111}$Cd & 1063 & 1429 & $^{149}$Sm & 1257 & 1885 & $^{189}$Os & 1509 & 2624 \\

$^{32}$S & 312 & 375 & $^{74}$Se & 667 & 902 & $^{112}$Cd & 1062 & 1455 & $^{150}$Sm & 1256 & 1903 & $^{190}$Os & 1507 & 2645 \\

$^{33}$S & 353 & 396 & $^{75}$As & 685 & 952 & $^{112}$Sn & 1115 & 1414 & $^{151}$Eu & 1264 & 1905 & $^{191}$Ir & 1515 & 2653 \\

$^{34}$S & 359 & 402 & $^{76}$Se & 666 & 957 & $^{113}$In & 1089 & 1461 & $^{152}$Sm & 1290 & 1938 & $^{192}$Os & 1533 & 2663 \\

$^{35}$Cl & 373 & 424 & $^{77}$Se & 701 & 984 & $^{114}$Cd & 1061 & 1450 & $^{153}$Eu & 1260 & 1941 & $^{192}$Pt & 1522 & 2661 \\

$^{36}$S & 368 & 441 & $^{78}$Se & 701 & 1011 & $^{114}$Sn & 1114 & 1467 & $^{154}$Sm & 1287 & 1979 & $^{193}$Ir & 1519 & 2671 \\

$^{36}$Ar & 386 & 430 & $^{78}$Kr & 699 & 965 & $^{115}$Sn & 1114 & 1465 & $^{154}$Gd & 1268 & 1926 & $^{194}$Pt & 1535 & 2679 \\

$^{37}$Cl & 390 & 451 & $^{79}$Br & 717 & 1016 & $^{116}$Sn & 1113 & 1463 & $^{155}$Gd & 1267 & 1946 & $^{195}$Pt & 1532 & 2688 \\

$^{38}$Ar & 397 & 460 & $^{80}$Se & 701 & 1064 & $^{117}$Sn & 1113 & 1478 & $^{156}$Gd & 1321 & 1966 & $^{196}$Pt & 1543 & 2697 \\

$^{39}$K & 410 & 481 & $^{80}$Kr & 733 & 1020 & $^{118}$Sn & 1112 & 1494 & $^{156}$Dy & 1309 & 1929 & $^{196}$Hg & 1533 & 2692 \\

$^{40}$Ar & 419 & 501 & $^{81}$Br & 717 & 1069 & $^{119}$Sn & 1111 & 1509 & $^{157}$Gd & 1319 & 1986 & $^{197}$Au & 1546 & 2704 \\

$^{40}$Ca & 423 & 486 & $^{82}$Kr & 733 & 1074 & $^{120}$Sn & 1110 & 1524 & $^{158}$Gd & 1318 & 2006 & $^{198}$Pt & 1539 & 2716 \\

$^{41}$K & 440 & 509 & $^{83}$Kr & 772 & 1101 & $^{120}$Te & 1147 & 1504 & $^{158}$Dy & 1306 & 1970 & $^{198}$Hg & 1529 & 2711 \\

$^{42}$Ca & 460 & 515 & $^{84}$Kr & 810 & 1127 & $^{121}$Sb & 1128 & 1530 & $^{159}$Tb & 1338 & 2008 & $^{199}$Hg & 1560 & 2720 \\

$^{43}$Ca & 466 & 535 & $^{84}$Sr & 782 & 1082 & $^{122}$Sn & 1108 & 1568 & $^{160}$Gd & 1315 & 2046 & $^{200}$Hg & 1558 & 2729 \\

$^{44}$Ca & 470 & 555 & $^{85}$Rb & 777 & 1132 & $^{122}$Te & 1146 & 1535 & $^{160}$Dy & 1357 & 2010 & $^{201}$Hg & 1551 & 2738 \\

$^{45}$Sc & 489 & 560 & $^{86}$Kr & 809 & 1179 & $^{123}$Sb & 1126 & 1573 & $^{161}$Dy & 1356 & 2029 & $^{202}$Hg & 1549 & 2747 \\

$^{46}$Ca & 476 & 593 & $^{86}$Sr & 782 & 1136 & $^{124}$Sn & 1106 & 1610 & $^{162}$Dy & 1354 & 2049 & $^{203}$Tl & 1556 & 2753 \\

$^{46}$Ti & 507 & 520 & $^{87}$Sr & 781 & 1163 & $^{124}$Te & 1144 & 1579 & $^{162}$Er & 1343 & 2059 & $^{204}$Hg & 1590 & 2765 \\

$^{47}$Ti & 511 & 548 & $^{88}$Sr & 838 & 1189 & $^{124}$Xe & 1181 & 1544 & $^{163}$Dy & 1353 & 2069 & $^{204}$Pb & 1551 & 2758 \\

$^{48}$Ti & 514 & 575 & $^{89}$Y & 786 & 1193 & $^{125}$Te & 1143 & 1600 & $^{164}$Dy & 1351 & 2088 & $^{205}$Tl & 1552 & 2771 \\

$^{49}$Ti & 517 & 602 & $^{90}$Zr & 792 & 1197 & $^{126}$Te & 1142 & 1621 & $^{164}$Er & 1393 & 2113 & $^{206}$Pb & 1559 & 2777 \\

$^{50}$Ti & 519 & 628 & $^{91}$Zr & 791 & 1206 & $^{126}$Xe & 1179 & 1587 & $^{165}$Ho & 1370 & 2154 & $^{207}$Pb & 1557 & 2786 \\

$^{50}$Cr & 550 & 593 & $^{92}$Zr & 848 & 1214 & $^{127}$I & 1159 & 1626 & $^{166}$Er & 1390 & 2167 & $^{208}$Pb & 1555 & 2795 \\

$^{51}$V & 537 & 638 & $^{92}$Mo & 831 & 1045 & $^{128}$Xe & 1177 & 1630 & $^{167}$Er & 1388 & 2193 &  \\

$^{52}$Cr & 555 & 648 & $^{93}$Nb & 867 & 1218 & $^{129}$Xe & 1175 & 1651 & $^{168}$Er & 1386 & 2220 &  \\

$^{53}$Cr & 557 & 653 & $^{94}$Zr & 847 & 1231 & $^{130}$Xe & 1174 & 1671 & $^{168}$Yb & 1428 & 2119 &  \\
\hline\hline
\end{tabular}
\end{table*}

\begin{table*}[ht]
\caption{\label{tab:kinenergieslonglived}Estimated bound kinetic energies (in MeV) of protons ($T^{p}_{\rm int}$) and neutrons ($T^{n}_{\rm int}$) in long-lived nuclides.}
\begin{tabular}{lrr|lrr|lrr|lrr|lrr}\hline\hline
Species & $T_{\rm int}^{p}$ & $T_{\rm int}^{n}$ &  Species & $T_{\rm int}^{p}$ & $T_{\rm int}^{n}$ & Species & $T_{\rm int}^{p}$ & $T_{\rm int}^{n}$ & Species & $T_{\rm int}^{p}$ & $T_{\rm int}^{n}$ & Species & $T_{\rm int}^{p}$ & $T_{\rm int}^{n}$ \\
$^{40}$K & 435 & 488 & $^{190}$Pt & 1526 & 2619 & $^{82}$Se & 780 & 1115 & $^{209}$Bi & 1573 & 2783 & $^{186}$Os & 1515 & 2560 \\

$^{87}$Rb & 824 & 1184 & $^{232}$Th & 1761 & 3048 & $^{96}$Zr & 846 & 1249 & $^{50}$V & 535 & 612 & $^{115}$In & 1087 & 1457 \\

$^{138}$La & 1218 & 1782 & $^{238}$U & 1804 & 3161 & $^{100}$Mo & 902 & 1186 & $^{113}$Cd & 1062 & 1453 & $^{123}$Te & 1145 & 1557 \\

$^{147}$Sm & 1260 & 1848 & $^{128}$Te & 1139 & 1663 & $^{116}$Cd & 1060 & 1480 & $^{144}$Nd & 1246 & 1844 & $^{152}$Gd & 1271 & 1889 \\

$^{176}$Lu & 1434 & 2325 & $^{76}$Ge & 668 & 999 & $^{130}$Te & 1136 & 1704 & $^{148}$Sm & 1259 & 1867 &  \\

$^{187}$Re & 1493 & 2593 & $^{48}$Ca & 480 & 629 & $^{150}$Nd & 1255 & 1950 & $^{174}$Hf & 1458 & 2247 &  \\
\hline\hline
\end{tabular}
\end{table*}

\end{document}